\documentclass[a4paper,11pt]{article}
\usepackage[latin1]{inputenc}
\usepackage[T1]{fontenc}
\usepackage[english]{babel}

\usepackage{graphicx}

\title{Generalised scalar-tensor theory in the Bianchi type I model}
\author{Stéphane Fay\\
66 route de la Montée Jaune\\
37510 Savonnières\\
\small{Steph.Fay@Wanadoo.fr}}
\date{10 March 1999}

\begin{document}
\maketitle
\begin{center}
Abstract
\end{center}
We use a conformal transformation to find solutions to the generalised scalar-tensor theory, with a coupling constant dependent on a scalar field, in an empty Bianchi type I model. We describe the dynamical behaviour of the metric functions for three different couplings: two exact solutions to the field equations and a qualitative one are found. They exhibit non-singular behaviours and kinetic inflation. Two of them admit both General Relativity and string theory in the low-energy limit as asymptotic cases.
\vspace{8cm}
\newline
Key words: Bianchi models; Generalised scalar-tensor theory; Exact solution; non-singular Universe; Kinetic inflation. 
\newpage
\section{Introduction} \label{s1}
Scalar-tensor theories seem to be essential to describe gravitational interactions near the Plank scale : string theory, higher order theories in the Ricci scalar \cite{4}, extended inflation and many others theories imply scalar field. 

The generalised scalar-tensor Lagrangian has the same form as the Brans-Dicke theory \cite{3} but with a coupling constant $\omega$ depending on the scalar field. Such a theory is interesting for many reasons. Hence, if we choose $\omega$ as a constant, the Lagrangian is identical to Brans-Dicke Lagrangian. This theory tends to General Relativity for large value of the coupling constant ($\omega>500$). But, if we choose $\omega=-1$, the Brans-Dicke theory is identical to the string theory in the low-energy limit. Hence, the generalised scalar-tensor theory seems to be able to build a " bridge " between string theory and General Relativity. Other reasons, as inflation, can be put forward : such a theory with a varying coupling constant, may drive the scale factors to accelerate without potential or cosmological constant \cite{6}\cite{7}, i.e. called kinetic inflation. 

The generalised scalar-tensor theory has been studied by many authors and the method we will use to find exact solutions has always been described in \cite{8} in the presence of matter in the Lagrangian. Here, we will consider the empty Bianchi type I Universe, which is spatially flat, and will use three different forms of the coupling constant $\omega(\phi)$. The first form, $2\omega(\phi)+3=2\beta(1-\phi/\phi_{c})^{-\alpha}$, has been introduced by Garcia-Bellido and Quiros \cite{1} and studied by Barrow \cite{2} in the context of a FLRW flat model with vacuum or radiation. It has also been studied in \cite{8}, for a Bianchi type I model, where a solution is found in presence of matter. In this paper, we will write explicitly an exact solution and will study the dynamical behaviour of the metric functions which depends on the integration constant. We will cast light on interesting features such as kinetic inflation. The second form is a power law type, $3+2\omega(\phi)=\phi_{c}^{2}\phi^{2m}$. Here again, we will give explicitly an exact solution and study it. An interesting feature is the possibility of a non-singular Universe. The third form is an exponential law type, $3+2\omega(\phi)=e^{2\phi_{c}\phi}$ and will be studied qualitatively. These two last laws seem interesting because power and exponential laws are very present in physics. They play a fundamental role for the metric functions of course, but also when we consider a potential $V(\phi)$ \cite{9} giving birth to extended or chaotic inflation \cite{10}. Moreover, we will see how the power law form of the coupling constant is linked to minimally coupled and induced gravity for large or small values of the scalar field.

This paper is organised as follows. In section \ref{s2}, we write field equations in both Brans-Dicke and conformal frame and  explain how to proceed to solve them. In section \ref{s3}, we derive solution for each of the three forms of $\omega(\phi)$ and study them.
\section{Field equations.} \label{s2}
\subsection{Field equations in the Brans-Dicke frame.} \label{s21}
We work with the metric:
\begin{equation} \label{1}
ds^{2}=-dt^{2}+a(t)^{2}(\omega^{1})^{2}+b(t)^{2}(\omega^{2})^{2}+c(t)^{2}(\omega^{3})^{2}
\end{equation}
$a(t)$, $b(t)$, $c(t)$ are the metric functions, $\omega^{i}$ are the 1-forms of the Bianchi type I model and $t$, the proper time. We express the Lagrangian of the theory in the form:
\begin{equation} \label{2a}
L=-\phi R+\omega(\phi)\phi_{,\alpha}\phi^{,\alpha}\phi^{-1}
\end{equation}
One can also cast (\ref{2a}) on the form:
\begin{equation} \label{2b}
L=-f(\Phi)R+1/2\partial_{\alpha}\Phi\partial^{\alpha}\Phi
\end{equation}
with
\begin{equation} \label{3}
\omega(\phi)=1/2ff^{,-2}
\end{equation}
The corresponding field equations and Klein-Gordon equation are obtained by varying the action (\ref{2a}) with respect to the space-time metric and the scalar field. If we introduce the $\tau$ time through 
\begin{equation} \label{4}
abcd\tau=dt
\end{equation}
then, denoting $d/d\tau$ by a prime, the field equations are:
\begin{eqnarray} 
\frac{a^{,,}}{a}-\frac{a^{,2}}{a^{2}}+\frac{a^{,}}{a}\frac{\phi^{,}}{\phi}-\frac{1}{2}\frac{\omega^{,}}{3+2\omega}\frac{\phi^{,}}{\phi}&=&0 \nonumber \\
\frac{b^{,,}}{b}-\frac{b^{,2}}{b^{2}}+\frac{b^{,}}{b}\frac{\phi^{,}}{\phi}-\frac{1}{2}\frac{\omega^{,}}{3+2\omega}\frac{\phi^{,}}{\phi}&=&0\\ \label{5}
\frac{c^{,,}}{c}-\frac{c^{,2}}{c^{2}}+\frac{c^{,}}{c}\frac{\phi^{,}}{\phi}-\frac{1}{2}\frac{\omega^{,}}{3+2\omega}\frac{\phi^{,}}{\phi}&=&0 \nonumber \\
\frac{a^{,}}{a}\frac{b^{,}}{b}+\frac{a^{,}}{a}\frac{c^{,}}{c}+\frac{b^{,}}{b}\frac{c^{,}}{c}+\frac{\phi^{,}}{\phi}(\frac{a^{,}}{a}+\frac{b^{,}}{b}+\frac{c^{,}}{c})-\frac{\omega}{2}(\frac{\phi^{,}}{\phi})^{2}&=&0 \label{6} \\
\phi^{,,}=-\frac{\omega^{,}\phi^{,}}{3+2\omega} \label{7}\\ \nonumber
\end{eqnarray} \nonumber
We can integrate (\ref{7}) to obtain the useful equation :
\begin{equation} \label{8}
A\phi^{,}\sqrt{3+2\omega}=1
\end{equation}
$A$ being an integration constant. Hence, we see that $\omega>-3/2$.
\subsection{Field equations in the conformal frame.} \label{s22}
Now, we work with the conformal metric:
\begin{equation} \label{11}
ds^{2}=-d\tilde{t}^{2}+\tilde{a}(t)^{2}(\omega^{1})^{2}+\tilde{b}(t)^{2}(\omega^{2})^{2}+\tilde{c}(t)^{2}(\omega^{3})^{2}
\end{equation}
By the conformal transformation the metric has been redefined as:
\begin{equation} \label{12}
\tilde{g}_{\alpha \beta}=\phi g_{\alpha \beta}
\end{equation}
and the Lagrangian becomes :
\begin{equation} \label{13}
L=R-1/2(3+2\omega)\phi_{,\alpha}\phi^{,\alpha}\phi^{-2}
\end{equation}
Hence, the generalised scalar-tensor theory is cast into Einstein gravity with a minimally coupled scalar field. In the $\tilde{\tau}$ time defined as :
\begin{equation} \label{14}
\tilde{a}\tilde{b}\tilde{c}d\tilde{\tau}=d\tilde{t}
\end{equation}
the field equations and the klein-Gordon equation become in the conformal frame
\begin{eqnarray} 
\frac{\tilde{a}^{,,}}{\tilde{a}^{,}}-\frac{\tilde{a}^{,}}{\tilde{a}}&=&0 \nonumber \\
\frac{\tilde{b}^{,,}}{\tilde{b}^{,}}-\frac{\tilde{b}^{,}}{\tilde{b}}&=&0 \label{15} \\
\frac{\tilde{c}^{,,}}{\tilde{c}^{,}}-\frac{\tilde{c}^{,}}{\tilde{c}}&=&0 \nonumber \\
\frac{\tilde{a}^{,}}{\tilde{a}}\frac{\tilde{b}^{,}}{\tilde{b}}+\frac{\tilde{a}^{,}}{\tilde{a}}\frac{\tilde{c}^{,}}{\tilde{c}}+\frac{\tilde{b}^{,}}{\tilde{b}}\frac{\tilde{c}^{,}}{\tilde{c}}&=&\frac{1}{2}(\omega+3/2)(\frac{\phi^{,}}{\phi})^{2} \label{16} \\
\frac{\phi^{,,}}{\phi^{,}}-\frac{\phi^{,}}{\phi}&=&-\frac{\omega^{,}}{3+2\omega}\label{17}\\ \nonumber
\end{eqnarray}
Equations (\ref{15}) are exactly the same as in the Bianchi type I model in General Relativity. Only the constraint equation (\ref{16}) is different. The solutions of the field equations are in the $\tilde{t}$ time the well-known Kasnerian solutions:
\begin{equation} \label{18}
\tilde{a}=\tilde{t}^{p_{1}}, \tilde{b}=\tilde{t}^{p_{2}}, \tilde{c}=\tilde{t}^{p_{3}} 
\end{equation}
$p_{1}$, $p_{2}$, $p_{3}$ being the Kasner exponents with :
\begin{equation} \label{19}
\sum p_{i}=1
\end{equation}
With the constraint equation, we obtain :
\begin{equation} \label{20}
\sum p_{i}^{2}=1-2\phi_{0}^{-2}
\end{equation}
$\phi_{0}$ being the integration constant of the scalar field. Hence, for all coupling constant $\omega(\phi)$, in the conformal frame, there will always be one negative Kasner exponent or three positive Kasner exponents and then two or three decreasing metric functions.
In the $\tilde{\tau}$ time, the solutions of (\ref{15}) are:
\begin{eqnarray}
\tilde{a}=e^{\alpha_{1}\tilde{\tau}+\alpha_{0}} \nonumber \\
\tilde{b}=e^{\beta_{1}\tilde{\tau}+\beta_{0}} \label{21} \\
\tilde{c}=e^{\gamma_{1}\tilde{\tau}+\gamma_{0}} \nonumber \\ \nonumber
\end{eqnarray}
where $\alpha_{i}$, $\beta_{i}$, $\gamma_{i}$ are integration constants. We integrate the Klein-Gordon equation to obtain the important equation:
\begin{equation} \label{22}
\tilde{\phi}_{0}\phi^{,}\phi^{-1}\sqrt{3+2\omega}=1
\end{equation}
$\tilde{\phi}_{0}$ being   an   integration   constant   (in  fact $\tilde{\phi}_{0}=A$). Hence,  we  deduce from the constraint equation that:
\begin{equation} \label{23}
\alpha_{1}\beta_{1}+\alpha_{1}\gamma_{1}+\beta_{1}\gamma_{1}=1/4\tilde{\phi}_{0}^{2}\mbox{,        }\forall\omega(\phi)
\end{equation}
To find solutions to the field equations (\ref{5}) we proceed as follow: first, we have to find solutions, for the scalar field, of the equations (\ref{8}) and (\ref{22}) so that we obtain respectively $\phi(\tau)$ and $\phi(\tilde{\tau})$. Second, we write $\phi(\tau)=\phi(\tilde{\tau})$ and reverse $\phi(\tilde{\tau})$ to find $\tilde{\tau}=\tilde{\tau}(\tau)$. Third, using (\ref{12}),  we write : 
\begin{equation} \label{24}
a=\tilde{a}(\tilde{\tau}(\tau))/\phi(\tau),b=\tilde{b}(\tilde{\tau}(\tau))/\phi(\tau),c=\tilde{c}(\tilde{\tau}(\tau))/\phi(\tau)
\end{equation}
Let us examine what are the relations between the quantities in the $\tau$ time and in the $t$ time. The amplitudes of the metric functions are the same in the both time since $a(\tau)=a(\tau(t))=a(t)$. The sign of the first derivatives are also the same : remember that $d\tau/dt=1/abc$ is positive since the metric functions are positive-definite. Hence, $\tau$ is an increasing function of $t$ and the sign of the first derivative of the metric functions will be the same in both $\tau$ time and $t$ time. The sign of the second derivatives in the $t$ time and $\tau$ time are different. If an overdot denotes differentiation with respect to $t$, the sign of $\ddot{a}$ will be that of $a^{,,}-a^{,}(a^{,}/a+b^{,}/b+c^{,}/c)$. We will study both the sign of $a^{,,}$ and $\ddot{a}$ in the applications of section \ref{s3}. Of course, the amplitudes of the derivatives are different in the $t$ and $\tau$ times. But we will not study them since we are mainly interested in their signs and therefore dynamical behaviour of the metric functions: whether they are increasing, decreasing or bouncing, and whether there is inflation.

Another difference between the two times comes from their asymptotic behaviours. For instance, the $t$ time could diverge at a finite value of the $\tau$ time. It depends mainly on $dt/d\tau=abc=V$, where $V$ is the volume of the Universe. In the cases we are going to study, the volume will always tend toward $0$ or infinity (we will show it for the two first theories of section \ref{s3}). Then, if $V\rightarrow 0$ when $\tau$ tends toward a constant or infinity, $t$ tends toward a constant. If $V\rightarrow \infty$ when $\tau\rightarrow \infty$, $t\rightarrow \infty$. If $V\rightarrow \infty$ when $\tau$ tends toward a constant, $t$ may tend toward infinity or a constant. In this last case, we need to integrate the volume $abc$ to make the asymptotic behaviour of  the cosmic time $t$ precise. Unhappily, it will not be possible in the theories of section \ref{s3}. We have studied the behaviour of the volume for the two first theories so that one can always get the asymptotic behaviour of $t(\tau)$ by using these rules (except the case $\tau\rightarrow cte$ and $V\rightarrow \infty$).

Concerning the presence of singularity, to ensure that a theory is non-singular, we will check that the Ricci curvature scalar $R$ is finite. The Ricci scalar can be writen:
\begin{equation} \label{24r}
R=(abc)^{-2}\left[-\omega(\phi^{,}\phi^{-1})^{2}+3\phi^{-1}\omega^{,}\phi^{,}(3+2\omega)^{-1}\right]
\end{equation}
\section{Non-singular and accelerated behaviours.} \label{s3}
To simplify the study of the metric functions, we will consider  in what follows only an increasing function of the scalar field, which means the only positive constants are $A$ and $\tilde{\phi}_{0}$.
\subsection{The case $3+2\omega=2\beta(1-\phi/\phi_{c})^{-\alpha}$} \label{s31}
We use the form for the coupling constant $3+2\omega=2\beta(1-\phi/\phi_{c})^{-\alpha}$ where $\beta$ is a positive constant, $\alpha$, $\phi_{c}$ are constant. The case $\alpha=0$ corresponds to Brans-Dicke theory and the case $\alpha=1$ and $\beta=-1/2$ to Barker's theory \cite{5}. Barrow showed in his paper \cite{2} that the case $\alpha=2$ is representative of the behaviour of other cases with $\alpha\not =2$ in the neighbourhood of the singularity. Hence, we will consider only this case. From (\ref{8}) we derive:
\begin{equation} \label{25}
\phi(\tau)=\phi_{c}\left[1-e^{-(\tau+\tau_{0})/(A\sqrt{2\beta}\phi_{c})}\right]
\end{equation}
from (\ref{22}) we deduce:
\begin{equation} \label{26}
\phi(\tilde{\tau})=\phi_{c}(1+e^{-(\tilde{\tau}_{0}+\tilde{\phi}_{0}^{-1}\tilde{\tau})/\sqrt{2\beta}})^{-1}
\end{equation}
Equating (\ref{25}) and (\ref{26}), we get:
\begin{equation} \label{27}
\tilde{\tau}=\tilde{\phi}_{0}\sqrt{2\beta}ln\left[e^{(\tau+\tau_{0})/(A\sqrt{2\beta}\phi_{c})}-1\right]-\tilde{\phi}_{0}\tilde{\tau}_{0}
\end{equation}
$\tau_{0}$ being an integration constant. Hence, using (\ref{24}), we write:
\begin{equation} \label{28}
a(\tau)=\frac{e^{-\tilde{\phi}_{0}\tilde{\tau}_{0}\alpha_{1}+\alpha_{0}}}{\sqrt{\phi_{c}}}(e^{\frac{\tau+\tau_{0}}{A\sqrt{2\beta}\phi_{c}}}-1)^{\sqrt{2\beta}\tilde{\phi}_{0}\alpha_{1}}(1-e^{-\frac{\tau+\tau_{0}}{A\sqrt{2\beta}\phi_{c}}})^{-1/2}
\end{equation}
and identical expressions for $b(\tau)$, $c(\tau)$ with $\beta_{0}$, $\beta_{1}$ and $\gamma_{0}$, $\gamma_{1}$ respectively. If we introduce:
\begin{equation} \label{29}
u=(\tau+\tau_{0})/(A\sqrt{2\beta}\phi_{c}),\mbox{ }a_{0}=e^{-\tilde{\phi_{0}}\tilde{\tau_{0}}\alpha_{1}+\alpha_{0}}/\sqrt{\phi_{c}}>0,\mbox{ }\alpha_{1}=-\sqrt{2\beta}\tilde{\phi}_{0}\alpha_{1}
\end{equation}
the expression (\ref{28}) becomes:
\begin{equation} \label{30}
a(\tau)=a_{0}(e^{u}-1)^{-a_{1}-1/2}e^{u/2}
\end{equation}
$u$ and the $\tau$ time vary in the same manner as long as $A$ and $\phi_{c}$ are positive constants. The constraint equation (\ref{23}) is rewritten as:
\begin{equation} \label{31}
a_{1}b_{1}+a_{1}c_{1}+b_{1}c_{1}=\frac{1}{2}\beta
\end{equation}
The metric function will be real for positive $u$. One can show that there is no non-singular behaviour for this theory in an anisotropic Universe. The Ricci curvature can be written as:
\begin{equation} \label{31a}
R=(e^{u}-1)^{1+2(a_{1}+b_{1}+c_{1})}(3-2\beta e^{2u}-24\beta^{2}e^{4u}+24\beta^{2}e^{5u})(2a_{0}^{2}b_{0}^{2}c_{0}^{2}e^{3u})^{-1}
\end{equation}
We check that conditions to get finite $R$ for asymptotic times $(u\rightarrow 0,u\rightarrow \infty)$ are not compatible: for $u\rightarrow 0$ we need $a_{1}+b_{1}+c_{1}>-1/2$ whereas for $u\rightarrow +\infty$, we need $a_{1}+b_{1}+c_{1}<-3/2$. So there is always a singularity for the Ricci curvature at small or/and large times.

The first derivative of (\ref{30}) shows that the metric function $a(\tau)$ will have a minimum for $u=-ln(-2a_{1})$ and $a_{1}\in \left]0,-1/2\right[$. For small $u$, we have $\phi\rightarrow 0$, $\omega\rightarrow \beta-3/2$ and:
\begin{equation} \label{32}
a\approx a_{0}(e^{u}-1)^{-a_{1}-1/2}
\end{equation}
Hence, if $a_{1}<-3/2$, $da/d\tau$ and $a$ tend to 0, if $a_{1}\in \left[ -3/2,-1/2\right]$, $da/d\tau$ tends to infinity and $a$ tends to 0, if $a_{1}>-1/2$, $da/d\tau$ and $a$ tends respectively to $-\infty$ and $+\infty$.
For large $u$, we have $\phi\rightarrow \phi_{c}$, $\omega\rightarrow +\infty$ if $\alpha>0$ and:
\begin{equation} \label{33}
a\approx a_{0}e^{-a_{1}u}
\end{equation}
Hence, if $a_{1}<0$, $da/d\tau$ and $a$ tend to infinity, if $a_{1}>0$, $da/d\tau$ and $a$ tend to 0. We see that the form of the metric function depends only on the parameter $a_{1}$:
\begin{itemize}
\item If $a_{1}<-3/2$, the metric function is increasing (fig 1).
\item If $a_{1}\in \left[-3/2,-1/2\right]$, it is increasing but with an inflexion point (fig 2). By studying the second derivative of $a(\tau)$, one can show that the condition to have an inflexion point is $a_{1}\in \left[-3/2,-1/2\right]$. In the other cases, the second derivative is always positive and the dynamic is always accelerated. Lets note that it is not inflation since for that we must have $\ddot{a}>0$ and not $a^{,,}>0$.
\item If $a_{1}\in \left[-1/2,0\right]$, the metric function has a minimum. Hence, if $a_{1}$, $b_{1}$ and $c_{1}$ belong to $\left[-1/2,0\right]$, all the metric functions have a bounce. However that does not mean that the Universe is non-singular since in this case the Ricci scalar become infinite for large $\tau$.
\item If $a_{1}>0$, the metric function is decreasing (fig 4).
\end{itemize}
Example of these four behaviours are illustrated on figures 1-4. 

Now we examine the sign of the second derivative of the metric function $a$ in the $t$ time so that we can detect inflation. It is the same as the quantity $-2a_{1}(b_{1}+c_{1})e^{2u}+(1-b_{1}-c_{1})e^{u}-1$ which is a second degree equation for $e^{u}$. One finds two roots: $e^{u_{1,2}}=(1-b_{1}-c_{1}\pm\sqrt{\Delta})\left[4a_{1}(b_{1}+c_{1})\right]^{-1}$ with $\Delta=(b_{1}+c_{1}-1)^{2}-8a_{1}(b_{1}+c_{1})$. If they are complex or inferior to 1, the sign of $\ddot{a}$ is the same as $-2a_{1}(b_{1}+c_{1})$. If they are superior to 1, there are two inflexion points: $\ddot{a}$ is first positive (negative), negative (positive) and then positive (negative) if $-2a_{1}(b_{1}+c_{1})>0$ ( respectively $-2a_{1}(b_{1}+c_{1})<0$). For the same reasons, if one of the roots is not real or inferior to 1, there is one inflexion point and $\ddot{a}$ is first positive (negative) and then negative (positive) if $-2a_{1}(b_{1}+c_{1})>0$ (respectively $-2a_{1}(b_{1}+c_{1})<0$). Here, $\ddot{a}>0$ can correspond to inflation when in the same time $\dot{a}$, or equivalently $a^{,}$, is positive. Hence, one see an example of kinetic inflation as described by Jana Levin in \cite{6} and \cite{7}. We remark also that inflation can end in a natural way.\\
\\
If now we write the volume:
\begin{equation} \label{34}
V=abc
\end{equation}
For small (large) $u$, $V$ vanishes if $a_{1}+b_{1}+c_{1}<-3/2$ ($a_{1}+b_{1}+c_{1}>0$) else it tends toward infinity.
Another interesting feature of this model is that for $\beta=1/2$, we have $\omega\rightarrow -1$ for small value of $u$, $\omega\rightarrow \infty$ and $\omega^{-3}(d\omega/d\phi)\rightarrow 0$ if $\alpha>1/2$ for large value. That is the two value of the coupling constant that corresponds to String theory in the low-energy limit and to General Relativity (by General Relativity we means that the post-Newtonian parameters of General Relativity are recovered).
\subsection{The case $3+2\omega=\phi_{c}^{2}\phi^{2m}$.} \label{s32}
Now, we consider the following form of the coupling constant:
\begin{equation} \label{35}
3+2\omega=\phi_{c}^{2}\phi^{2m}
\end{equation}
where $\phi_{c}$ and $m$ are real constants. Using the same process than before, from (\ref{8}) we derive :
\begin{equation} \label{36}
\phi (\tau)=\left[(m+1)/(A\phi_{c})(\tau+\tau_{0})\right]^{1/(m+1)}
\end{equation}
and from (\ref{22}) we get :
\begin{equation} \label{37}
\phi (\tilde{\tau})=\left[m(\tilde{\phi}_{0}\phi_{c})^{-1}(\tilde{\tau}+\tilde{\tau}_{0})\right]^{1/m}
\end{equation}
Equating (\ref{36}) and (\ref{37}) we have:
\begin{equation} \label{38}
\tilde{\tau}=\frac{\tilde{\phi}_{0}\phi_{c}}{m}\left[\frac{m+1}{A\phi_{c}}(\tau+\tau_{0})\right]^{m/(m+1)}-\tilde{\tau}_{0}
\end{equation}
Then, with (\ref{24}) we obtain :
\begin{equation} \label{39}
a(\tau)=exp\{\frac{\alpha_{1}\tilde{\phi}_{0}\phi_{c}}{m}\left[\frac{m+1}{A\phi_{c}}(\tau+\tau_{0})\right]^{m/(m+1)}-\alpha_{1}\tilde{\tau}_{0}+\alpha_{0}\}\left[\frac{m+1}{A\phi_{c}}(\tau+\tau_{0})\right]^{-1/2(m+1)}
\end{equation}
We introduce the variables :
\begin{equation} \label{40}
a_{0}=e^{-\alpha_{1}\tilde{\tau}_{0}+\alpha_{0}},\mbox{ }a_{1}=\alpha_{1}\tilde{\phi}_{0}\phi_{c},\mbox{ }u=(\tau+\tau_{0})/(A\phi_{c})
\end{equation}
and (\ref{39}) becomes :
\begin{equation} \label{41}
a=a_{0}exp(a_{1}m^{-1}\left[(m+1)u\right]^{m/(m+1)})\left[(m+1)u\right]^{-1/2(m+1)}
\end{equation}
We get the same type of expressions for $b(\tau)$ and $c(\tau)$. From the constraint equation (\ref{23}) we deduce:
\begin{equation} \label{42}
a_{1}b_{1}+a_{1}c_{1}+b_{1}c_{1}=\phi_{c}^{2}/4
\end{equation}
The expression (\ref{41}) of the metric function shows that $(m+1)u$ must be positive. Hence, if $m>-1$, $u\in \left[0,+\infty\right[$ and if $m<-1$, $u\in \left]-\infty,0\right]$. $u$ and the $\tau$ time vary in the same manner as long as $A$ and $\phi_{c}=\sqrt{\phi_{c}^{2}}$ are two positive constants.

First, let us examine the Ricci scalar. It is written:
\begin{eqnarray} \label{42r}
R=\left[(1+m)u\right]^{(1-2m)/(1+m)}\mbox{[}3-\phi_{c}^{2} \left[(m+1)u\right]^{2m/(m+1)}+ \nonumber\\
6m\phi_{c}^{4} \left[(m+1)u\right]^{4m/(1+m)}  \mbox{]}\left[2a_{0}^{2}b_{0}^{2}c_{0}^{2}(1+m)^{2}e^{2(a_{1}+b_{1}+c_{1}) \left[(m+1)u\right]^{m/(m+1)}/m}\right]^{-1}\\ \nonumber
\end{eqnarray}
Only if $m\in\left[0,1/2\right]$ and $a_{1}+b_{1}+c_{1}>0$, is the Ricci scalar always finite at both small and large times, avoiding the singularity. Now we examine the dynamic of $a$ in the $\tau$ time. The first derivative of (\ref{41}) vanishes for $u=(2a_{1})^{-(m+1)/m}/(m+1)$ and hence, $a(\tau)$ has an extremum for this value that exists only if $a_{1}$ is positive. The asymptotic study of (\ref{41}) when $u\rightarrow 0$ and $u\rightarrow \pm\infty$ gives the results summarised in table 1. We found eight different behaviours. The figures 5-12 show an example of each of them. To summarise the main characteristics of each case in the $\tau$ time:
\begin{itemize}
\item For $a_{1}<0$, the metric function is always decreasing and has an inflexion point when $m<-3/2$.
\item For $a_{1}>0$, the metric function has a minimum if $m>0$ and a maximum if $m<0$. Hence, only the case where $a_{1},b_{1},c_{1}$ and $m$ are positive, gives birth to a "bounce" Universe. It avoids the singularity if $m\in\left[0,1/2\right]$ and $a_{1}+b_{1}+c_{1}>0$ and will be today in expansion in all directions of space. 
\end{itemize}
If we define the volume $V$ by $V=abc$, then it tends to vanish for small $u$ if $m/(m+1)<0$ and $m( a_{1}+b_{1}+c_{1} )<0$ or $m/(m+1)>0$ and $-3/\left[2(m+1)\right]>0$. It becomes infinite if $m/(m+1)<0$ and $m( a_{1}+b_{1}+c_{1})>0$ or $m/(m+1)>0$ and $-3/\left[2(m+1)\right]<0$. For large $u$, it tends to vanish if $m/(m+1)>0$ and $m( a_{1}+b_{1}+c_{1})<0$ or $m/(m+1)<0$ and $-3/\left[2(m+1)\right]<0$. It becomes infinite if $m/(m+1)>0$ and $m( a_{1}+b_{1}+c_{1})>0$ or $m/(m+1)<0$ and $-3/\left[2(m+1)\right]>0$.\\
\\
By examining the sign of $a^{,,}$, we can conclude that the dynamic of the metric function will always be accelerated (recall again that it is not inflation since it does not mean that $\ddot{a}>0$) if $m>1/2$ or $m\in\left[-3/2,1/2\right]$ and $a_{1}<0$. If $m<-3/2$ the dynamic is first accelerated and then decelerated. The same thing happens when $m\in\left[0,1/2\right]$ and $a_{1}>0$ whereas for $m\in\left[-3/2,0\right]$ and $a_{1}>0$, the metric accelerates again.\\
\\
We complete this study by examining the sign of the second derivative in the $t$ time. It is the same as $m+(b_{1}+c_{1})\left[ \left[(m+1)u\right]^{m/(m+1)}-2a_{1}\left[(m+1)u\right]^{2m/(m+1)}\right]$. This is a second degree equation for $\left[(m+1)u\right]^{m/(m+1)}$. The two roots are
\begin{equation}
u_{1,2}=(m+1)^{-1}\left[(b_{1}+c_{1}\pm\sqrt{\Delta})(4a_{1}(b_{1}+c_{1}))^{-1}\right]^{(m+1)/m}\nonumber
\end{equation}
with $\Delta=(b_{1}+c_{1})(8a_{1}m+b_{1}+c_{1})$. If $u_{1,2}$ are not real, the sign of $\ddot{a}$ is the one of $-2a_{1}(b_{1}+c_{1})$. When the two roots are real, they always belong to the interval where $u$ varies since their sign is the same as $m+1$. Then $\ddot{a}$ has the same sign as $-2a_{1}(b_{1}+c_{1})$ if $u$ is out of $\left[u_{1},u_{2}\right]$ or the opposite sign if $u\in\left[u_{1},u_{2}\right]$. There are two inflexion points. Hence, we get the same type of behaviour for $\ddot{a}$ as in the previous subsection. In the same manner, if only one root is real, the dynamic of $a$ will be accelerated and then decelerated or vice-versa depending on the sign of $-2a_{1}(b_{1}+c_{1})$. So, there is one inflexion point. For this theory also, inflation can end naturally.\\
\\
Concerning the coupling constant, we have for $m+1>0$: when $\tau\rightarrow +\infty$, $\phi\rightarrow +\infty$, $\omega\rightarrow \phi_{c}^{2}\phi^{2m}/2\rightarrow +\infty$  if $m>0$ and $\omega\rightarrow -3/2$ if $m\in\left[-1,0\right]$. When $\tau\rightarrow \tau_{0}$, $\phi\rightarrow 0$, $\omega\rightarrow \phi_{c}^{2}\phi^{2m}/2\rightarrow +\infty$  if $m\in\left[-1,0\right]$ and $\omega\rightarrow -3/2$ if $m>0$. Considering these last remarks and the relation (3), one can deduce that the asymptotic behaviours of the metric functions when $\phi\rightarrow 0$, $\omega\rightarrow \phi_{c}^{2}\phi^{2m}/2\rightarrow +\infty$ and $m\in\left[-1,0\right]$ are the same as in the cases of a coupling function of type $f(\Phi)=f_{0}e^{n\Phi}$ when $\phi_{c}^{2}=n^{-2}$ and $m=-1/2$ and $f(\Phi)=(f_{0}\Phi+f_{1})^{n}$ when $\phi_{c}^{2}=(f_{0}n)^{-2}$ and $2m=(2-n)/n$ with $n\not\in\left[0,2\right]$. Moreover, the asymptotic behaviour of the metric functions when $\phi\rightarrow +\infty$, $\omega\rightarrow \phi_{c}^{2}\phi^{2m}/2\rightarrow +\infty$ and $m>0$ are the same as in the previous case but with $n\in\left[0,2\right]$.\\
Hence the study of the metric functions when $3+2\omega=\phi_{c}^{2}\phi^{2m}$, give us information on the asymptotic behaviours of two different couplings $f(\Phi)$, that is $f(\Phi)=(f_{0}\Phi+f_{1})^{n}$ and $f(\Phi)=f_{0}e^{n\Phi}$. For the first of these functions, the minimally coupled theory is obtained for $f_{0}=0$ and $f_{1}^{n}=1/2$ , whereas the induced gravity is obtained for $f_{1}=0$, $f_{0}=\sqrt{\epsilon/2}$ and $n=2$. We note that the study of one coupling constant $\omega(\phi)$ permit us to get informations on two types of coupling $f(\Phi)$ because $\omega(\phi)$ and $f(\Phi)$ are linked by the differential equation (\ref{3}). Hence to one type of function $\omega$, having one or several free parameters, can correspond more than one type of functions $f$. What we say above comes from the fact that to a power or exponential law for $f(\Phi)$ correspond only a power law for $\omega(\phi)$.
\subsection{The case $3+2\omega=e^{2\phi_{c}\phi}$.} \label{s33}
We take the form $3+2\omega=e^{2\phi_{c}\phi}$, $\phi_{c}$ being a real constant. This is an interesting case because, as in the subsection \ref{s31}, when the scalar field vanishes, the coupling constant tends towards -1, which is the low limit of the string theory, whereas when it becomes infinite, the coupling constant tends towards infinity and the theory towards General Relativity if $\phi_{c}>0$.

Here, we can not integrate equation (\ref{22}) in a closed convenient form. We rewrite the equations (\ref{8}) and (\ref{22}) in the following form:
\begin{equation} \label{43}
H(\phi)=\tau=\int Ae^{\phi_{c}\phi}d\phi-\tau_{0}=A\phi_{c}^{-1}e^{\phi_{c}\phi}-\tau_{0}
\end{equation}
\begin{equation} \label{44}
G(\phi)=\tilde{\tau}=\int \tilde{\phi}_{0}e^{\phi_{c}\phi}\phi^{-1}d\phi-\tilde{\tau}_{0}
\end{equation}
That means we have $\phi(\tau)=H^{(-1)}(\tau)$ and $\phi(\tilde{\tau})=G^{(-1)}(\tilde{\tau})$. By equalling these last two expressions and reversing (\ref{43}), we get :
\begin{equation} \label{45}
\tilde{\tau}=G(H^{(-1)})=G(\phi)=G(\phi_{c}^{-1}ln\left[\phi_{c}A^{-1}(\tau+\tau_{0})\right])
\end{equation}
With (\ref{24}), we can easily obtain the metric functions :
\begin{equation} \label{46}
a=e^{\alpha_{1}G(\phi_{c}^{-1}ln\left[\phi_{c}A^{-1}(\tau+\tau_{0})\right])+\alpha_{0}}/\sqrt{(A\phi_{c})^{-1}ln\left[\phi_{c}(\tau+\tau_{0})\right]}
\end{equation}
and the same form for $b(\tau)$ and $c(\tau)$ with their integration constants. The reality conditions for the metric functions will be $\phi_{c}A^{-1}(\tau+\tau_{0})>0$ and $\phi_{c}^{-1}ln\left[\phi_{c}A^{-1}(\tau+\tau_{0})\right]>0$.

Hence, if $\phi_{c}<0$, the metric function will be real if $\tau\in\left]A\phi_{c}^{-1}-\tau_{0},-\tau_{0}\right[$, and if $\phi_{c}>0$, we will have $\tau\in\left]A\phi_{c}-\tau_{0},+\infty\right[$. The first derivative of (\ref{46}) will be of the sign of $\alpha_{1}\tilde{\phi}_{0}\phi_{c}A^{-1}(\tau+\tau_{0})-1/2$. For all value of $\phi_{c}$, when $\tau=A(2\alpha_{1}\phi_{c}\tilde{\phi}_{0})^{-1}-\tau_{0}$, $da/d\tau$ vanishes in the following cases:

- when $\tau\in\left]A\phi_{c}^{-1}-\tau_{0},-\tau_{0}\right[$, that means $\phi_{c}<0$, if $2\alpha_{1}\tilde{\phi}_{0}>1$,

- when $\tau\in\left]A\phi_{c}^{-1}-\tau_{0},+\infty\right[$, that means $\phi_{c}>0$, if $2\alpha_{1}\tilde{\phi}_{0}\in\left[0,1\right]$.

From these results and after a numerical study we can write that :
\begin{itemize}
\item If $\phi_{c}<0$, $\tau\in\left]A\phi_{c}-\tau_{0},-\tau_{0}\right[$:
	\begin{itemize}
	\item If $\alpha_{1}<(2\tilde{\phi}_{0})^{-1}$, the metric function is decreasing and tends to infinity, in a positive manner when $\tau\rightarrow A\phi_{c}^{-1}-\tau_{0}$, and to zero when $\tau\rightarrow -\tau_{0}$.
	\item If $\alpha_{1}>(2\tilde{\phi}_{0})^{-1}$, the metric function tends to zero for these two values of $\tau$ and has a maximum . So, if the three integration constants $\alpha_{1}$, $\beta_{1}$, $\gamma_{1}$ of each of the metric functions are such that they are all superior to $(2\tilde{\phi}_{0})^{-1}$, we have a close Universe (for the time) which exists during a finite time in the $\tau$-time. Since $dt/d\tau=abc$, this quantity vanishes in $\tau=A\phi_{c}^{-1}-\tau_{0}$ and $\tau=\tau_{0}$ and then $t(\tau)$ stays finite for these two values and the Universe also exists during a finite t time.
	\end{itemize}
\item If $\phi_{c}>0$, $\tau\in\left]A\phi_{c}-\tau_{0},+\infty\right[$ : 
	\begin{itemize}
	\item If $\alpha_{1}<0$, the metric function decreases from infinity to zero. 
	\item If $\alpha_{1}\in\left[0,(2\tilde{\phi}_{0})^{-1}\right]$, the metric function has a minimum and tends to $+\infty$ when $\tau$ tends to $A\phi_{c}^{-1}-\tau_{0}$ or $+\infty$. If the three integration constants $\alpha_{1}$, $\beta_{1}$, $\gamma_{1}$ are all in the same interval, the Universe will have a bounce since each metric function has a minimum. 
	\item If $\alpha_{1}>(2\tilde{\phi}_{0})^{-1}$, the metric function is increasing from zero to infinity with an infinite slope.
	\end{itemize}
\end{itemize}
\section{Conclusion} \label{s4}
In the conformal frame, the scalar field is minimally coupled. Hence, the spatial components of the field equations are exactly the same as in General Relativity and their solutions for the Bianchi type I model are the kasnerian solutions \cite{8}. The Klein-Gordon equation and the constraint equation, that are different from General Relativity, impose that the sum of the square of the Kasner exponents is always inferior to unity. Their sum is equal to one. Hence, there are always two or three positive Kasner exponents.

To express the metric function in the Brans-Dicke frame, we have equated the expressions of the scalar field in both Brans-Dicke and conformal frames and then deduced the time $\tilde{\tau}$ of the conformal frame as a function of the time $\tau$ of the Brans-Dicke frame. Then it is easy to find  the form of the metric functions in this last frame. The amplitude of the metric functions and the sign of their first derivative in the $\tau$ time of the Brans-Dicke frame are the same as in the $t$  time. This is not the case for the second derivative of the metric functions.

We have studied three forms of the coupling constant $\omega(\phi)$ and found solutions for which the Universe could avoid the singularity. We have also detected kinetic inflation for the two first examples and notice that, under some conditions, it can end naturally. For small or large value of the $\tau$ time, the coupling constant can become infinite or constant. It is always interesting to find classes of coupling constant for which it tends naturally toward -1 or infinite for small or large  value of $\tau$ because such a class of theories tends respectively toward string theory in the low-energy limit and General Relativity. It seems to be true in the special case $3+2\omega=(1-\phi/\phi_{c})^{-2}$ and for $3+2\omega=e^{2\phi_{c}\phi}$.

\newpage
\begin{table}
\begin{center}
\begin{tabular}{|c|c|c|}
\hline
 & $a_{1}<0$ & $a_{1}>0$ \\
\hline
m>0 & $u\rightarrow 0^{+},a\rightarrow +\infty,a^{,}\rightarrow -\infty$ &$u\rightarrow 0^{+},a\rightarrow +\infty,a^{,}\rightarrow -\infty$ \\
 & $u\rightarrow +\infty,a\rightarrow 0^{+},a^{,}\rightarrow 0$ & $u\rightarrow +\infty,a\rightarrow +\infty,a^{,}\rightarrow +\infty$ \\
\hline
$m\in\left[-1,0\right]$ & $u\rightarrow 0^{+},a\rightarrow +\infty,a^{,}\rightarrow -\infty$ & $u\rightarrow 0^{+},a\rightarrow 0,a^{,}\rightarrow +\infty$ \\
 & $u\rightarrow +\infty,a\rightarrow 0^{+},a^{,}\rightarrow 0$ & $u\rightarrow +\infty,a\rightarrow 0,a^{,}\rightarrow 0$ \\
\hline
$m\in\left[-3/2,-1\right]$ & $u\rightarrow 0^{-},a\rightarrow 0^{+},a^{,}\rightarrow 0$ &$u\rightarrow 0^{-},a\rightarrow 0,a^{,}\rightarrow -\infty$ \\
 & $u\rightarrow -\infty,a\rightarrow +\infty,a^{,}\rightarrow -\infty$ & $u\rightarrow -\infty,a\rightarrow 0,a^{,}\rightarrow 0$ \\
\hline
$m<-3/2$ & $u\rightarrow 0^{-},a\rightarrow 0,a^{,}\rightarrow 0$ &$u\rightarrow 0^{-},a\rightarrow 0,a^{,}\rightarrow -\infty$ \\
 & $u\rightarrow -\infty,a\rightarrow +\infty,a^{,}\rightarrow -\infty$ & $u\rightarrow -\infty,a\rightarrow 0,a^{,}\rightarrow 0$ \\
\hline
\end{tabular}
\caption{The eight different asymptotic behaviours of the metric function when $3+2\omega=\phi_{c}^{2}\phi^{2m}$. The asymptotic amplitudes of $a$ are the same in $t$ and $\tau$ time. That is not the case for the amplitudes of the first derivatives. We do not examine the asymptotic behaviour of the amplitudes of $\dot{a}$ since we are mainly interested by the study of the exact solutions in the $\tau$ time and, in a general maner, by the signs of $a^{,}$, $a^{,,}$ and $\ddot{a}$. But this is always possible by calculating $\dot{a}=a^{,}(abc)^{-1}$.}
\end{center}
\end{table}
\newpage
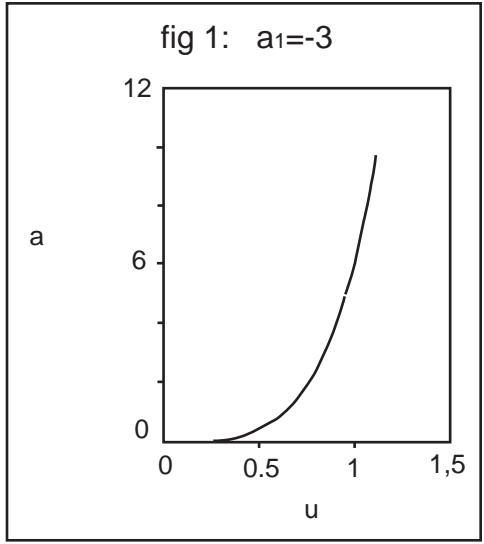
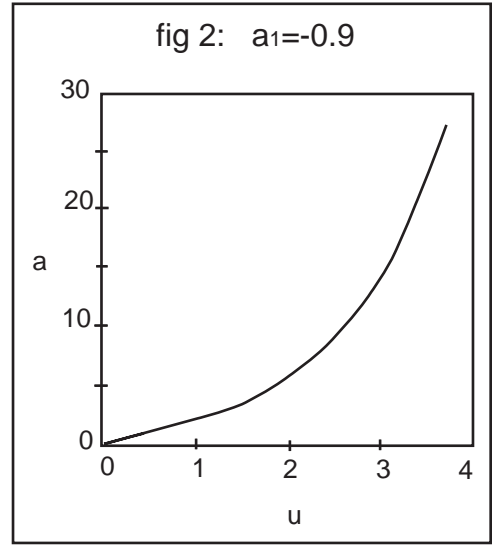
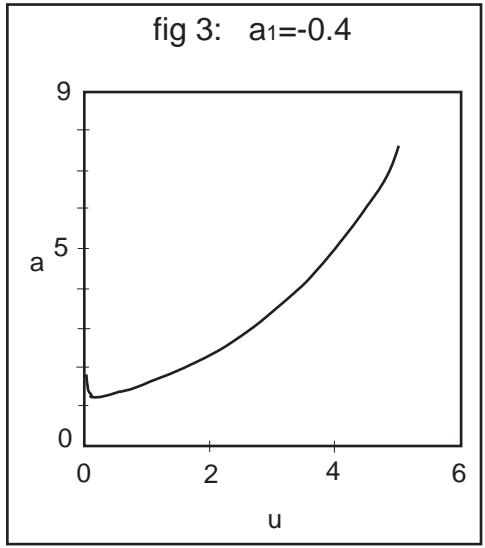
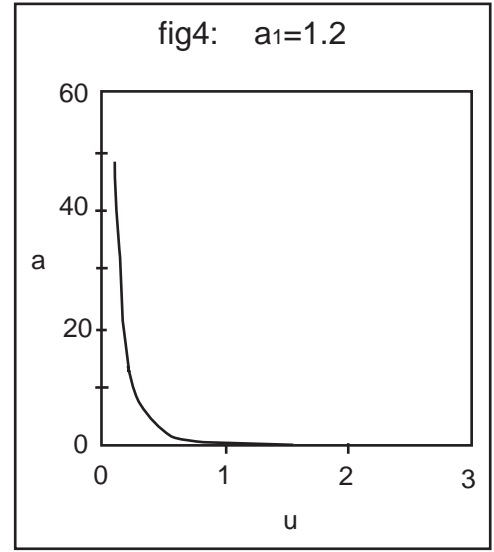
Figures 1 to 4 : forms of the metric functions when $3+2\omega=2\beta(1-\phi/\phi_{c})^{-2}$.\\
\\
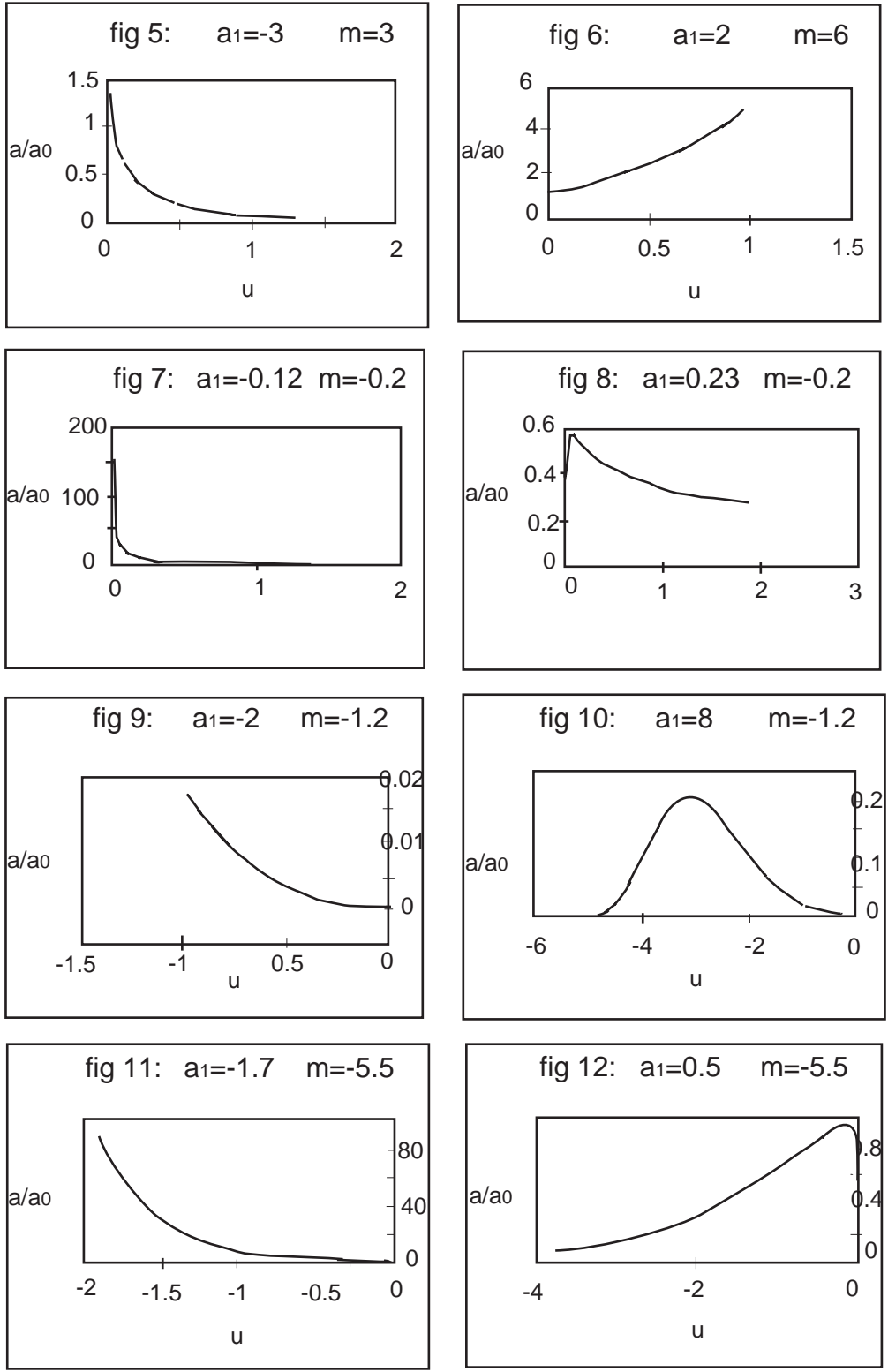
Figures 5 to 12 : forms of the metric functions when $3+2\omega=\phi_{c}^{2}\phi^{2m}$.


\begin{thebibliography}{9}

\bibitem{4}
D. Wands,
\emph{Extended gravity theories and the Einstein-Hilbert action},
Class. Quant. Grav 11, 269,
(1994).

\bibitem{3}
C. Brans and RH. Dicke,
Phys Rev 124, 925,
(1961).

\bibitem{6}
Janna Levin,
\emph{Kinetic inflation in stringy and other cosmologies},
Phys Rev D51,
pp 1536,
(1995).

\bibitem{7}
Janna Levin,
\emph{Gravity-Driven acceleration of the cosmic expansion},
Phys Rev D51,
pp 462,
(1995).

\bibitem{8}
J.P. Mimoso, D. Wands,
\emph{Anisotropic scalar-tensor cosmologies},
Phys Review D,
vol 52, p5612-5627,
(1995).

\bibitem{1}
J. Garcia-Bellido and M Quiros,
Phys. Lett. B, 
243, 45 ,
(1990).

\bibitem{2}
John D. Barrow,
Phys Review D,
Vol 48, n°8,
(1993).

\bibitem{9}
P.Parson, J.D. Barrow,
\emph{Generalised scalar field potentials and inflations},
Phys. Rev D51, 6757-6763,
(1995).

\bibitem{10}
E.W.Kolb,
\emph{First-Order inflation},
Physica Scripta,
Vol T36, 199-217,
(1991).

\bibitem{5}
BM. Barker,
Astro Phys J, 219, 5
(1978).

\end{thebibliography}
\end{document}